\input{aipcheck}

\documentclass[
    ,final            
  ]
  {aipproc}

\layoutstyle{6x9}

\def\be{\begin{equation}}
\def\ee{\end{equation}}
\def\bea{\begin{eqnarray}}
\def\eea{\end{eqnarray}}

\def\Z1{\widetilde{Z}_1}

\begin{document}

\title{Numerical Study of the Ghost-Ghost-Gluon Vertex on the Lattice}

\author{A.\ Mihara, A.\ Cucchieri and T.\ Mendes}{
address={Instituto de F\'{\i}sica de S\~ao Carlos, Universidade
de S\~ao Paulo,  \\C.P. 369, 13560-970, S\~ao Carlos, SP, Brazil}
}

\begin{abstract}
It is well known that, in Landau gauge, the renormalization function
of the ghost-ghost-gluon vertex $\Z1(p^2)$ is finite and
constant, at least to all orders of perturbation theory.
On the other hand, a direct non-perturbative verification of this result
using numerical simulations of lattice QCD is still missing.
Here we present a preliminary numerical study of the ghost-ghost-gluon
vertex and of its corresponding renormalization function
using Monte Carlo simulations in $SU(2)$ lattice Landau gauge.
Data were obtained in 4 dimensions for lattice couplings $\beta = 2.2, \,
2.3, \, 2.4$ and lattice sides $N = 4,\, 8,\, 16$. 
\end{abstract}

\maketitle


\section{Introduction}

Although the Faddeev-Popov ghosts are absent from the physical spectrum
of QCD, 
one can use the ghost-ghost-gluon vertex and the ghost
propagator to calculate physical observables such as the
strong running coupling \cite{Bloch:2003sk}
\be
\alpha_s(p^2) \; = \; \alpha_0\,
  \frac{Z_3(p^2) \, \widetilde{Z}_3^2(p^2)}{\Z1^2(p^2)} \; .
\label{eq:alpha_run}
\ee
Here $\alpha_0 = g_0^2 / 4 \pi $ is the bare coupling constant and
$Z_3(p^2)$, $\widetilde{Z}_3(p^2)$ and $\Z1(p^2)$ are, respectively,
the gluon, ghost and ghost-ghost-gluon vertex renormalization functions.

In Landau gauge the renormalization function
of the ghost-ghost-gluon vertex $\Z1(p^2)$ is finite and
constant, i.e.\ independent of the renormalization scale $\mu$,
at least to all orders of perturbation theory
\cite{Taylor:ff}. This implies that in this gauge one can consider
a definition of the running coupling constant
(in the momentum subtraction scheme) given by \cite{vonSmekal:1997is}
\be
\alpha_s(p^2) \; = \; \alpha_0\, p^6\;D(p^2)\, G^2(p^2) \; ,
\label{eq:alpha_run2}
\ee
where $\,D(p^2)\,$ and $\,G(p^2)\,$ are, respectively, the
gluon and ghost propagators.
An indirect evaluation of the ghost-ghost-gluon renormalization function
on the lattice
has been recently presented in \cite{Bloch:2003sk}, confirming
that $\Z1(p^2)$ is finite in the continuum limit.
On the other hand, a direct non-perturbative verification of this
result using lattice simulations is still missing.
Let us stress that a direct evaluation of $\Z1(p^2)$
would allow \cite{Furui:2000mq}
a study of the running coupling constant
using eq.\ (\ref{eq:alpha_run}) instead of eq.\ (\ref{eq:alpha_run2}).
Such a study may improve the precision of the determination of 
$\alpha_s(p^2)$, since in this case one will not need to
use a {\em matching rescaling} technique \cite{Bloch:2003sk}
when considering data evaluated at different lattice couplings $\beta$.
Moreover, a direct study of $\Z1(p^2)$ is interesting to determine its
$\beta$-dependence and
possible effects due to the breaking of rotational symmetry.
Here we present a preliminary numerical study of the
Landau-gauge ghost-ghost-gluon vertex
and of its corresponding renormalization function $\Z1(p^2)$
in the pure $SU(2)$ case.

\section{The vertex on the lattice}

The 3-point function (in momentum space) for ghost,
anti-ghost and $A_{\mu}^a$ (gluon) fields is
\be
V^{abc}_{\mu}(k;q,s) \; = \;
 (2 \pi)^4
   \; \delta^4(k+q-s) \;\langle \, A^{a}_{\mu}(k) \, G^{bc}(q) \, \rangle 
                              \,\, ,
\ee
where $G^{bc}(y,z)$ is the inverse of the Faddeev-Popov matrix
and $\delta^4(k+q-s)$ corresponds to momentum conservation.
Then, we can define the ghost-ghost-gluon vertex
function by ``amputating''  the corresponding 3-point function, i.e.\
\be
\Gamma^{abc}_{\!\mu}(k,q) \; = \;
\frac{V \; \langle \, A^{a}_{\mu}(k) \, G^{bc}(q) \, \rangle}{
         D(k^2) \, G(q^2) \, G(s^2)} \,\, ,
\ee
where V is the lattice volume and $s = k+q$.
From the weak-coupling expansion
in $SU(N_c)$ lattice gauge theory one obtains that   
the ghost-ghost-gluon vertex, at tree level, is
given by \cite{Kawai:1980ja}
\be
\Gamma_{\mu}^{abc}(k,q) \; =\; i \, g_0 \, f^{abc} \hat{q}_{\mu}
\cos\left(\frac{s_{\mu} a}{2}\right)\,\, ,
\label{eq:lamb0}
\ee
where $a$ is the lattice spacing.
Let us recall that a generic lattice momentum
$\hat{k}$ has components
$\hat{k}_{\mu} = 2\sin(k_{\mu}a/2) / a$, where
$k_{\mu} = 2\pi n_{\mu}/N_{\mu}$ and $n_{\mu}$, $N_{\mu}$ are,
respectively, the
Matsubara modes and the number
of lattice points in the $\mu$ direction.
Clearly, by taking the formal continuum limit $\,a \to 0\,$
in (\ref{eq:lamb0}) one recovers
the continuum tree-level result
$ \,\Gamma_{\mu}^{abc}(k,q)\; =\; i\, g_0 \, f^{abc} q_{\mu} \,$.
Finally, we obtain that the vertex renormalization function
in momentum-subtraction scheme is given by
\be
 \Z1^{-1}(p^2) \; = \;
    \frac{-i}{g_0 \, N_c  \,(N_c^2 \, - \, 1)} \;
 \sum_{\mu}\;
     \frac{\hat{p}_{\mu}}{\hat{p}^2\,\cos(p_{\mu}/2)}\,
              \sum_{a,b,c}\, f^{abc}\;\Gamma^{abc}_{\!\mu}(0,p) \,\, ,
\ee
where $N_c$ is the number of colors and
we consider the asymmetric point
with zero momentum for the gluon ($k = 0$), implying $s = q$.

\section{Numerical simulations and results}

\begin{figure}[t]
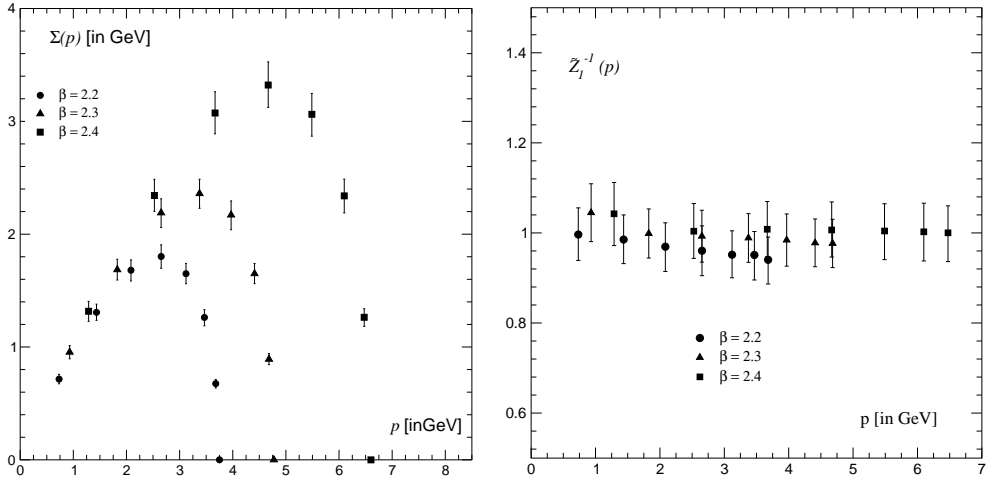

\includegraphics[width=0.42\textwidth]{Sym_Sigma.eps}
\protect\hspace{2mm}
\includegraphics[width=0.435\textwidth]{Sym_Z1.eps}
\caption{Results for
$\,\Sigma(p^2)\,$ and $\,\Z1^{-1}(p^2)\,$  with lattice volume $V = 16^4$.
For the lattice spacing in physical units we refer to
\protect\cite{Bloch:2003sk}. Error bars
were obtained using the bootstrap method with 250 samples.}
\label{fig:res2}
\end{figure}

For each lattice volume $V = N^4 = 4^4,\, 8^4,\,16^4\,$
we have performed simulations for three lattice couplings
($\beta = 2.2,\, 2.3,\,2.4\,$). We consider lattice momenta
with components
$\,\hat{p}_1 = \hat{p}_2 = \hat{p}_3 = \hat{p}_4 = \hat{p}\,$
and $\,p = 2\pi n/N$, where $\,n\,$ gets values $\,1, \ldots, N/2\,$.
With this choice of momenta we obtain
\bea
 \Z1^{-1}(p^2) \! \! \! & = &  \! \! \!
     \frac{1}{\hat{p}\,\cos(p/2)}\;\Sigma(p^2) \\
 \Sigma(p^2)   \! \! \!& = &\! \! \!
     \frac{1}{2\, g_0\,N_c  \,(N_c^2 \, - \, 1)} \;
      \sum_{a,b,c}\, f^{abc}\;
              \sum_{\mu}\;  
      \mbox{Im}\; \Gamma^{abc}_{\!\mu}(0,p) \,\, ,
\eea
where we used the fact that only the imaginary part of
the vertex function $\Gamma^{abc}_{\!\mu}(0,p)$ contributes to $\Z1^{-1}(p^2)$.
Results for the quantity $\Sigma(p^2)$ defined above
and the renormalization function
$\Z1(p^2)$ are presented in Fig.\ \ref{fig:res2}. Clearly,
$\Sigma(p^2)$ has the same momentum dependence as the tree-level
vertex [i.e.\ $\sim\hat{p}\cos(p/2)$], implying that $\Z1(p^2)$
is approximately constant with respect to the momentum $p$.
Also the dependence on $\beta$ seems to be very weak.
More details will be presented elsewhere \cite{preparation}.

\begin{theacknowledgments}
This work was supported by
FAPESP through grants 00/05047-5 (AC, TM) and 03/00928-1 (AM).
Partial support from CNPq is also acknowledged (AC, TM).
\end{theacknowledgments}


\bibliographystyle{aipproc}   

\bibliography{my_references}

\IfFileExists{\jobname.bbl}{}
 {\typeout{}
  \typeout{******************************************}
  \typeout{** Please run "bibtex \jobname" to optain}
  \typeout{** the bibliography and then re-run LaTeX}
  \typeout{** twice to fix the references!}
  \typeout{******************************************}
  \typeout{}
 }

\end{document}